# Interfacial strain defines the self-organization of epitaxial MoO$_2$ flakes and porous films on sapphire: experiments and modelling

O. de Melo,[1,2,3] V. Torres-Costa,[2] Y. González,[2,4] A. Ruediger,[4] C. de Melo,[5] J. Ghanbaja,[5] D. Horwat,[5] A. Escobosa,[6] O. Concepción,[2,6*] G. Santana,[3] E. Ramos[3]

[1]Physics Faculty, University of Havana, 10400 La Habana, Cuba

[2]Departamento de Física Aplicada, Universidad Autónoma de Madrid. Cantoblanco 28049, Madrid, Spain

[3]Instituto de Investigaciones en Materiales, Universidad Nacional Autónoma de México, Cd. Universitaria, A.P. 70-360, Coyoacán 04510, México D. F.

[4]Institut National de la recherche scientifique, Centre Énergie, Matériaux, Télécommunications, 1650 Boulevard Lionel-Boulet, Varennes, Québec, J3X 1S2, Canada

[5]Institut Jean Lamour, Université de Lorraine, UMR 7198, Nancy, F-54000, France

[6]Sección de Electrónica del Estado Sólido, Departamento de Ingeniería Eléctrica, CINVESTAV, IPN, Mexico

Abstract. The epitaxy of MoO$_2$ on c-plane sapphire substrates is examined. A theoretical approach, based on density functional theory calculations of the strain energy, allowed to predict the preferred layer/substrate epitaxial relationships. To test the results of these calculations, $MoO_2$/(001) $Al_2O_3$ heterostructures were grown using the chemically-driven isothermal close space vapour transport technique. At the early stages of the growth, two kinds of morphologies were obtained, using the same growth parameters: lying and standing flakes. The composition and morphology, as well as the layer/substrate epitaxial relationships were determined for both kind of morphologies. Experimental epitaxial relationships coincide with those predicted by DFT calculation as the most favourable ones in terms of strain energy. For thicker films, the standing flakes evolve to form an epitaxial porous layer composed by coalesced epitaxial flakes. The interfacial strain between the sapphire substrate and $MoO_2$ enables a self-organization from nanometer to micron scales between separated or coalesced flakes, depending on deposition condition.

---

* Presently at: MESA+ Institute for Nanotechnology, University of Twente, 7500 AE Enschede, the Netherlands.

1. Introduction

Molybdenum oxides are the subject of intense research efforts because of their interesting properties and promising applications in different areas. They cover catalysis, energy storage and harvesting, computing, and biomedical applications[1]. In many cases, the crystal surface plays a decisive role, reason why the study of molybdenum oxide flakes has become a subject of great interest. In fact, it has been demonstrated that the surface of $MoO_2$ flakes exhibits a very good performance for Surface Enhanced Raman Spectroscopy in terms of sensitivity, reusability, uniformity, and stability[2]. In particular, standing flakes (sometimes called vertical flakes) are especially interesting objects for applications requiring a large developed surface area. For example, vertical flakes of graphene are very promising due to the large surface to volume ratio with accessible channel between flakes and high mechanical stability, among other advantages[3]. On the other hand, to favour the transport of lithium to the anode active centres in lithium ion batteries, the use of porous $MoO_2$-based composites has been proposed [4]. Finally, prediction and control of the orientation of the flakes facets is relevant for catalysis applications because catalytic activity was been found to depend on the exposed crystalline planes of the catalyst[5].

The epitaxy of $MoO_2$ onto (001)-oriented $Al_2O_3$ (c-plane sapphire) is an interesting problem. As described in Table I, $Al_2O_3$ has a trigonal structure, frequently described by a hexagonal lattice, while $MoO_2$ presents a monoclinic structure that resembles a distorted hexagonal one. The values of the *a* and *c* lattice parameters of this monoclinic structure are similar and the angle between them ($\theta_{ac}$) is very close to 120°, as between *a* and *b* axes in $Al_2O_3$. Yet, those parameters are quite larger than the *a* and *b* lattice parameters of $Al_2O_3$. Based on these characteristics, it is not obvious to predict the epitaxial relation (if any) for the $MoO_2$/(001) $Al_2O_3$ heterostructures.

Synthesis of epitaxial thin films of $MoO_2$ by sputtering[6] and of epitaxial $MoS_2$@$MoO_2$ nanorods[7] using chemical vapour deposition at atmospheric pressure, with sulphur as reducing agent, have been reported recently. In those papers it was supposed that the film-substrate epitaxial relationships can be determined using interfacial strain considerations. Following qualitative considerations based on the lattice mismatch along different crystalline directions, the authors explained the obtained epitaxial relationships. Nevertheless, the correlation between growth pattern and supporting surface cannot be unambiguously determined with this qualitative approach.

In the present work, we carry out density functional theory (DFT) calculations of the strain energy of the elemental lattice cell of $MoO_2$ for different expected film orientations. This allowed to predict the preferred epitaxial relationships according to strain criteria on a quantitative basis. The results of this calculations were tested through the growth of $MoO_2$/(001) $Al_2O_3$ heterostructures with different growth parameters. For the growth experiments we used a technique reported in a previous work[8]: the chemically-driven isothermal close space vapour transport (CD-ICSVT). This technique allows to prepare pure $MoO_2$ films under a flow of reductive $H_2$ atmosphere and with $MoO_3$ as source material.

The CD-ICSVT growth of $MoO_2$ self-organized flakes at nanometer to micron scales (in dependence of the growth conditions) onto (001)-$Al_2O_3$ substrates is presented. X-ray diffraction, high-resolution scanning and transmission electron microscopy, transmission electron diffraction as well as optical microscopy and Raman spectroscopy, allowed to determine the crystalline orientation of the deposited flakes facets. Different epitaxial relationships with the substrate are observed for lying and standing flakes. Such relations prevail in conditions in which thicker porous epitaxial films are obtained by coalescence of isolated flakes. The obtained epitaxial relations are compared with the results of DFT calculations.

Table I. Parameters for the crystal lattices of $MoO_2$ and $Al_2O_3$

|  | Structure/space group | a (Å) | b (Å) | c (Å) | θ (°) |
|---|---|---|---|---|---|
| $MoO_2$ | Monoclinic/P1 $2_1$/c1 (14) | 5.584 | 4.842 | 5.608 | $\theta_{ac}$=120.983; $\theta_{ab}$= 90; $\theta_{bc}$= 90 |
| $Al_2O_3$ | Trigonal/R$\bar{3}$c (167) | 4.761 | 4.761 | 12.994 | $\theta_{ab}$=120; $\theta_{ac}$=90; $\theta_{bc}$=90 |

2. Methods.

The CD-ICSVT experimental setup and procedure have been described in details elsewhere[3]. A graphite boat containing the $MoO_3$ powder (99.5 % purity, provided by Sigma-Aldrich) with the substrate placed above, is located in the high-temperature plateau region of a tubular furnace. The system is flowed with a $H_2$:Ar (1:5) reductive gas mixture at a rate of 25 mL/min and at atmospheric pressure. Source to substrate distance ($d_{s-s}$) of around 6 mm was used while the growth temperature was varied between 570 and 705 °C and the growth time between 10 and 60 min. These conditions allowed obtaining the desired kinds of morphology: isolated flakes or relatively compact flaked thin films.

Calculations were carried out under the DFT framework[9,10], using the DMol3 code[11] implemented in the Materials Studio 8.0 software suite[12], employing periodical bonding conditions and the M06-L functional[13], and a double numeric basis set (DND), with an effective potential all electron relativistic. The occupied convergence tolerances for calculations were $1\times10^{-5}$ Ha, $2\times10^{-5}$ Ha, $4\times10^{-3}$ Ha/Å, and $5\times10^{-3}$ Å for SFC cycles, energy, gradient, and displacement, respectively. $MoO_2$ was modelled using the $MoO_2$ Crystallographic Information File (CIF 9009090) as an initial structure for the optimization process. The $MoO_2$ structure was fully optimized (including cell parameters). All other calculations were realized as a single point energy.

X-rays diffractograms were obtained in a θ-2θ configuration with an AXS Bruker D8 Advance diffractometer or, for pole figures and high-resolution measurements, with a Panalytical X'Pert Pro MRD diffractometer. In both cases, $K_\alpha Cu$ radiation (λ = 1.54 Å) was used. High resolution scanning electron microscopy (HR-SEM) images were obtained by using a FE-SEM Hitachi S-4700 microscope. Transmission electron microscopy (TEM) measurements were performed by a JEOL ARM 200-Cold FEG (point resolution 0.19 nm). Raman spectroscopy was carried out with a confocal optical microscope coupled with a modular Raman spectrometer from Horiba (iHR320). The laser (473 nm, Cobolt Inc.) focused onto the sample with a 100 × objective and a power of 2.79 mW was used.

3. Density functional theory calculations of the strain energy

The strain energy on the $MoO_2$ lattice, constrained by in-plane lattice vectors and angles of the $Al_2O_3$ surface, was evaluated for four different epitaxial configurations as described in Fig. 1 (oxygen atoms positions in the (001) sapphire surface are indicated with red circles). In this figure, relevant in-plane crystallographic directions are indicated by arrows (blue for sapphire and green, yellow or black for [001], [100] and [010], respectively, for MoO$_2$). These configurations were selected after a qualitative analysis which took into account strain induced by mismatches on both inter-planar distances and angles; the selected configurations have lower strain energy than others not considered in this work. (See the supplementary material, S1) for two other possible configurations which were tested and demonstrated to have much higher strain energy than the four selected ones). The strain energy for each configuration was determined by subtracting the calculated energy of the distorted crystal lattice (according a given epitaxial relation) from that of the un-strained MoO$_2$ crystal lattice.

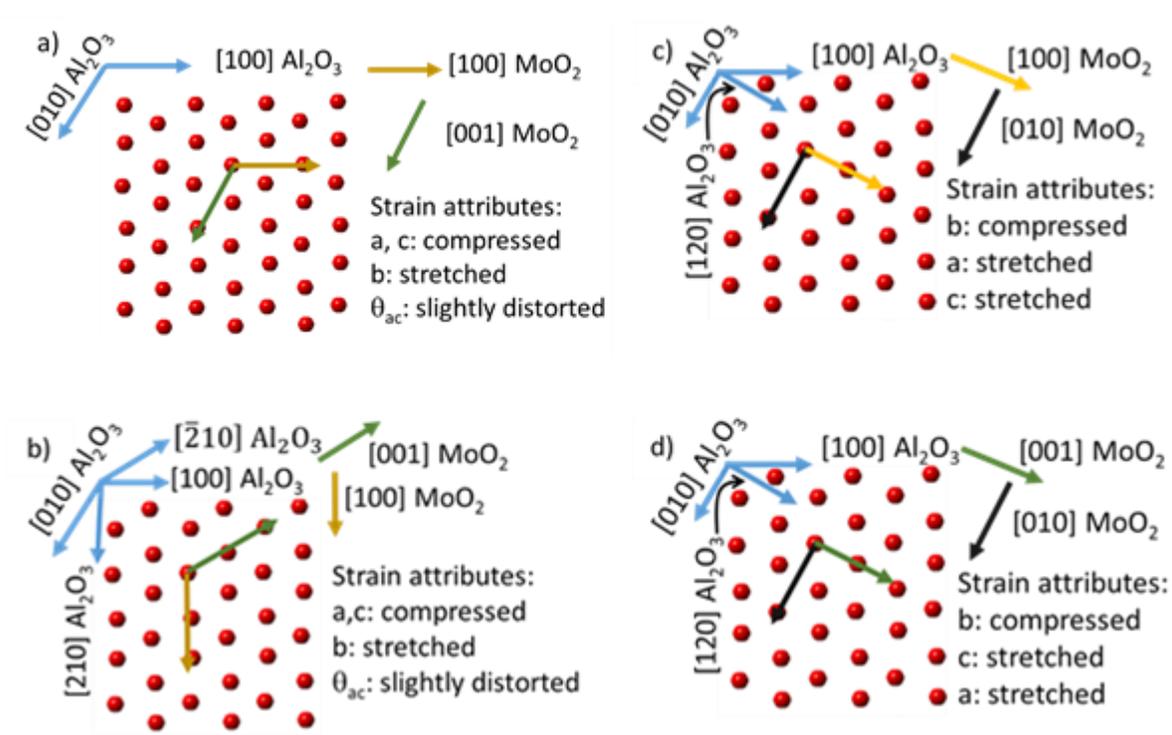

**Fig. 1. Epitaxial relationships, considered for strain energy calculations, between (001) Al$_2$O$_3$ surface (oxygen atoms positions indicated with red circles) and MoO$_2$ with different orientations. Oxygen atoms positions in the (001) sapphire surface are indicated with red circles while relevant in-plane crystallographic directions are indicated by arrows (blue for sapphire and green, yellow or black for**

**[001], [100] and [010], respectively, for MoO$_2$. a) Configuration A: $[001]\,\text{MoO}_2 \parallel [010]\,\text{Al}_2\text{O}_3$ and $[010]\,\text{MoO}_2 \parallel [001]\,\text{Al}_2\text{O}_3$; b) configuration B: $[100]\,\text{MoO}_2 \parallel [210]\,\text{Al}_2\text{O}_3$ and $[010]\,\text{MoO}_2 \parallel [001]\,\text{Al}_2\text{O}_3$; c) configuration C: $[100]\,\text{MoO}_2 \parallel [120]\,\text{Al}_2\text{O}_3$ and $[010]\,\text{MoO}_2 \parallel [010]\,\text{Al}_2\text{O}_3$; d) configuration D: $[001]\,\text{MoO}_2 \parallel [120]\,\text{Al}_2\text{O}_3$ and $[010]\,\text{MoO}_2 \parallel [010]\,\text{Al}_2\text{O}_3$. Configuration A and B correspond with the (010) plane parallel to the surface while configurations C and D with planes (001) and (100) parallel to the surface, respectively.**

Each of the configurations considered in Fig. 1 presents strain attributes which are qualitatively described in the figure. In every case, the two in-plane lattice parameters of MoO$_2$ (and the angle between them) are constrained to the substrate surface values. In the calculations, the third lattice parameter (out-of-plane) was left to relax in order to obtain its equilibrium value (the minimum energy of the strained cell). In configurations A and B (Fig. 1 a and b), MoO$_2$ would grow with the (010) crystalline plane parallel to the surface of the Al$_2$O$_3$ substrate while in configurations C and D (Fig. 1 c and d), the parallel planes would be the (001) and the (100), respectively. The results from calculations are shown in Fig. 2 in which the strain energy is plotted versus the out-of-plane lattice parameter in every configuration (*b* for configurations A and B; *a* for configuration C and *c* for configuration D).

The minimum of the curves shown in Fig. 2 indicates the equilibrium energy of the strained cells and the equilibrium out-of-plane lattice parameter. The obtained out-of-plane parameters values were 0.591, 0.510, 0.593 and 0.578 nm for configurations A to D, respectively. It is interesting to note that, according to these strain energy considerations, configuration A in which *a* and *c* lattice parameters of MoO$_2$ are parallel to *a* and *b* parameters of Al$_2$O$_3$, respectively, is significantly less stable than the other three configurations.

However, the other configurations present similar strain energy costs. If the epitaxial relationships of MoO$_2$ on sapphire is guided by strain energy considerations, the above results suggest that MoO$_2$ flakes grown on (001) Al$_2$O$_3$ could present more than one orientation. In the following sections, the experimental results about the growth of MoO$_2$ onto (001) Al$_2$O$_3$ will be contrasted with this DFT modelling.

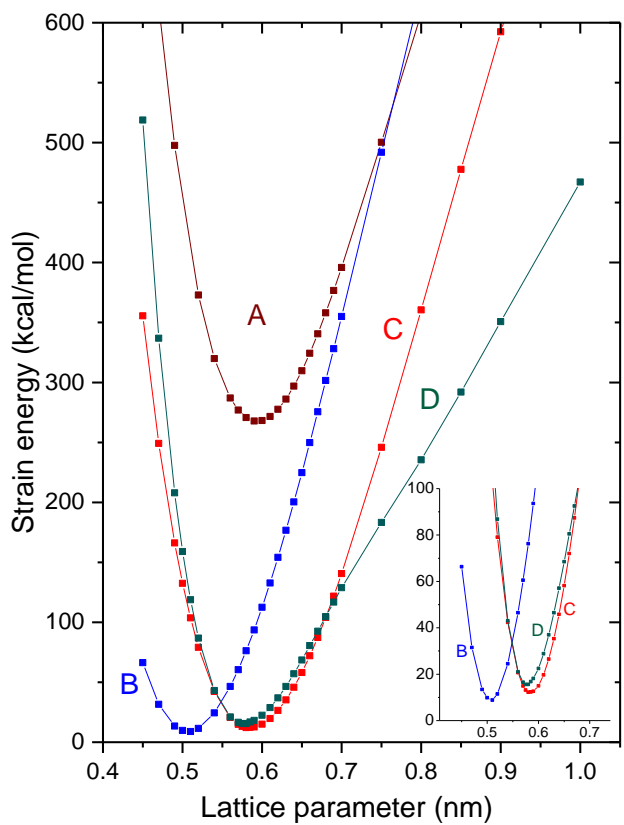

Fig. 2. Strain energy as a function of the corresponding out-of-plane lattice parameter (*b* for configurations A and B; *c* for configuration C and *a* for configuration D). As can be observed, the most stable configuration is B, competing with C and D. According to these strain energy arguments, the occurrence of configuration A should be discarded. The inset is a magnification of the lower strain energy region showing the values of the strain energies of the three most stable configurations.

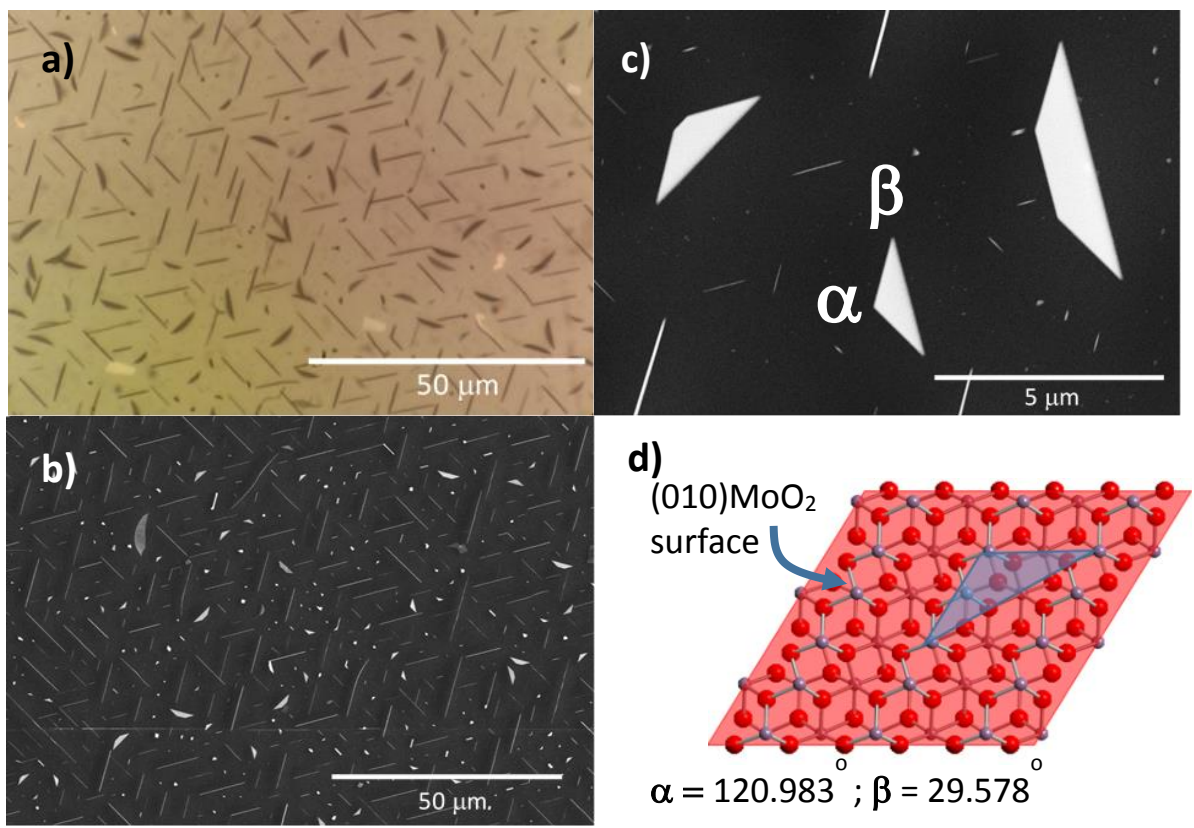

**Fig. 3.** Optical (a) and scanning electron microscopy (b, c) images of MoO$_2$ flakes grown onto c-cut sapphire substrate at two different magnifications. Two kinds of microscopic flakes can be observed: triangular ones (truncated or not) lying onto the sapphire surface, and standing elongated walls. It can be noted that the flakes are oriented along three directions which form 120° with each other. The vertices of the triangles are always oriented along one of the directions along which the largest dimension of the walls accommodates. In (d), a schematic of the (010) surface of MoO$_2$ is displayed (O and Mo atoms are represented by red and blue circles, respectively).

4. Structure and morphology of MoO$_2$ grown onto c-plane sapphire surfaces.

The appearance of the MoO$_2$ material grown onto the Al$_2$O$_3$ surface is shown in Fig. 3 at different scales. This sample was grown at 700 °C during 15 min. With these conditions, we were able to obtain repeatedly large isolated flakes. This can be appreciated by the optical (a) and scanning electron microscopy (b, c) images. In particular, two kinds of microscopic flakes are observed:

triangular ones (truncated or not) lying onto the sapphire surface, and standing elongated wall-type ones. It can be noted that the flakes are oriented along three directions which form 120° with each other. This is clearly evidenced in the Fourier transform of the image in b (available in the supplementary material, S2). The vertices of the triangles are always oriented along one of the directions along which the largest dimension of the walls accommodates. In Fig. 3d, a schematic of the (010) surface of $MoO_2$ is displayed (O and Mo atoms are represented by red and purple circles). Geometrical considerations suggest that laying flakes have (010) planes parallel to the (001) $Al_2O_3$ surface (A or B configurations).

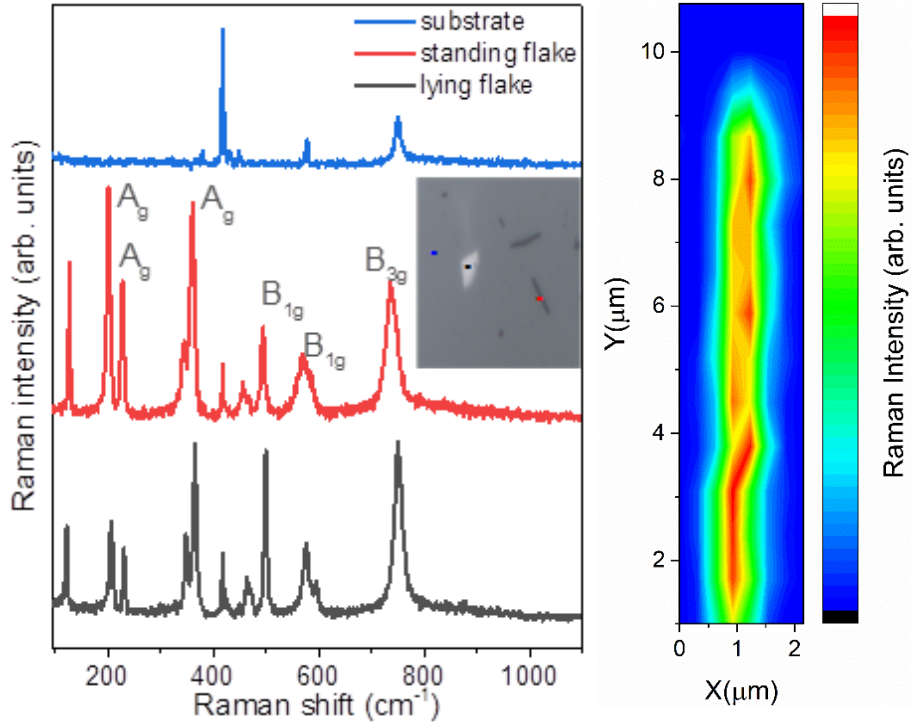

**Fig. 4.** Micro-Raman spectra in three different positions onto the surface of the sample: lying (black) and standing (red) flakes and the sapphire substrate surface (blue). $MoO_2$ peaks are identified according to the irreducible representation notation. At the right, a Raman map of the intensity of the peak at 128 cm$^{-1}$ around an elongated flake.

To support that flakes are composed by pure $MoO_2$, micro-Raman spectra taken in three different positions onto the sample surface are presented in Fig. 4 together with a map of the intensity of the peak at 128 cm$^{-1}$ around a standing flake. The spectra taken on the flakes show the characteristic peaks of $MoO_2$[14], which are identified in the figure through the irreducible representation notation. The rest of the surface is observed to be uncovered since the obtained Raman spectrum in this region corresponds with that of sapphire.

To study the crystalline orientation of these flakes, a diffractogram in the $\theta$-$2\theta$ configuration was carried out, which is shown in Fig. 5. Besides the reflections of the (001)-oriented $Al_2O_3$ substrate, only a few other peaks corresponding to $MoO_2$ were observed: two peaks around $2\theta = 37°$ and two less intense ones at about $2\theta = 79°$. It can reasonably be considered that these preferential orientations originate from the two different kinds of flakes observed in Fig. 3. According the geometrical analysis in Fig. 1d, and the expected peak positions for the $MoO_2$ (P121/c1 or group 14 system, PDF number 01-073-1249), we assigned one of the reflections at near 37° to the (020) plane of $MoO_2$ and one of the reflections at near 79° to the parallel plane (040), as indicated in the figure. These reflections correspond to the triangular lying flakes. Then, the other reflections coming from $MoO_2$ were assigned to the (002) and (004) crystalline planes of $MoO_2$ and correspond to the standing wall-like flakes. This assignment is supported by pole figure analysis of thicker films, as will be described below.

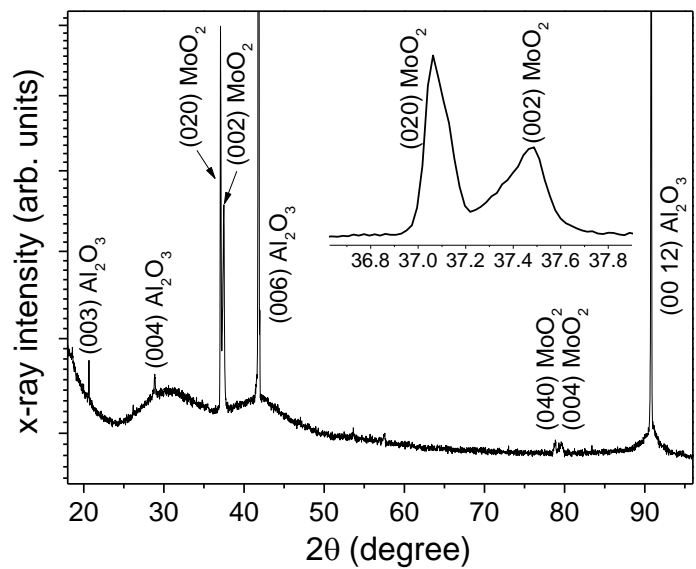

**Fig. 5.** X-ray diffractogram of a MoO$_2$ sample grown at 700 °C for 15 min onto (001) sapphire (c-cut). A strong preferential orientation is indicated by the absence of the (011) reflection which is expected to be the most intense one for randomly-oriented crystallites. Only two kinds of reflections are observed for MoO$_2$, corresponding to two different crystalline planes. The magnified 2θ range around 37° is presented as inset.

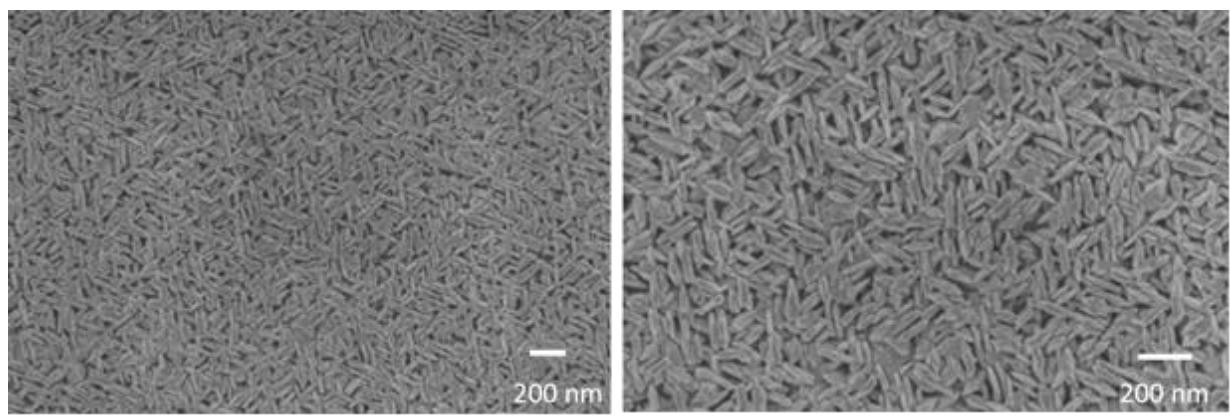

**Fig. 6.** SEM images of the surface of a sample grown at 600 °C for 1 h at two different magnifications. Standing walls present a compact arrangement in which the largest dimensions are oriented along three different directions, suggesting epitaxial growth on the substrate.

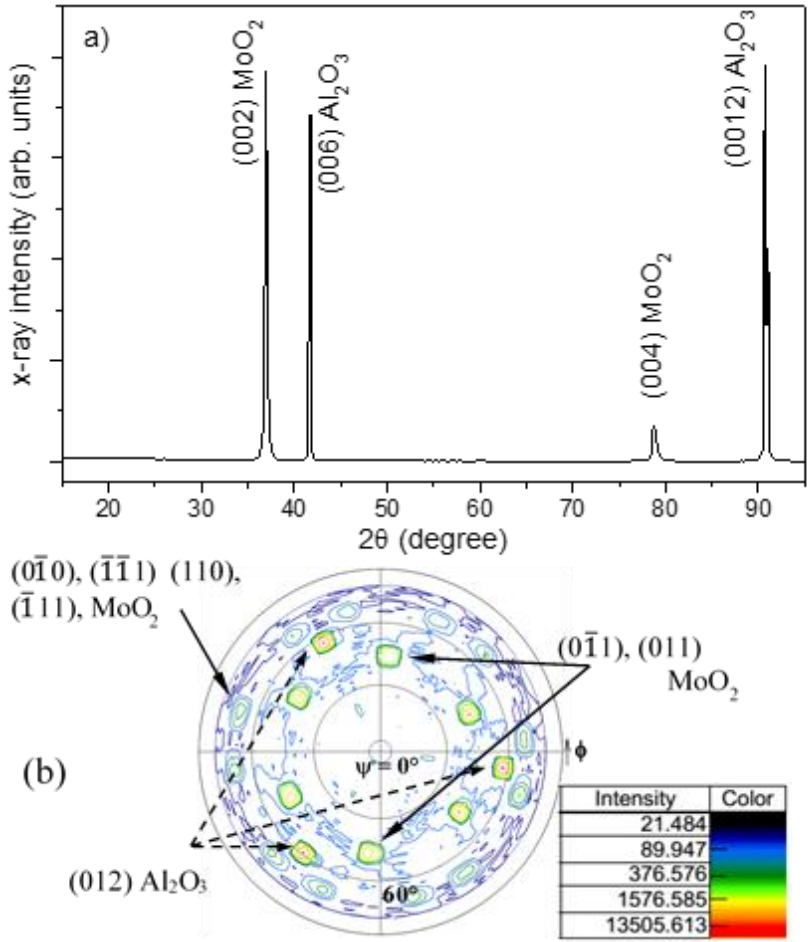

**Fig. 7.** (a) X-ray diffractogram of monoclinic MoO$_2$ structure onto Al$_2$O$_3$ substrate. Miller indices of the different crystallographic planes (hkl) are indicated. (b) Pole figure taken around a 2θ angle of 26°.

To analyse the behaviour of compact thin films instead of isolated flakes, we grew samples using growth parameters optimized in previous experiments: the temperature was decreased to 600 °C and the duration increased to 1 h. In these conditions, favouring nucleation and using larger times, we observe the growth of coalesced standing flakes of smaller sizes (Fig. 6). It can be noted that, as in the case of isolated flakes, the larger dimension of the flakes is oriented along three directions separated 120° from each other. A diffractogram of this sample is shown in Fig. 7a. Strong preferential orientation is observed as indicated by the intense peaks near 37 and 79°, as in the case of the isolated flakes. The presence of $MoO_2$ was corroborated by Raman measurements (not shown here).

Pole diagrams were carried out to determine the crystalline orientation of the standing flakes with respect to the substrate. A Bragg $2\theta$ angle of 26° was selected for this study because, very near to this angle, reflections from three different crystalline plane families of $MoO_2$ ({011}, {110} and {11$\bar{1}$}) together with the {012} plane family reflection of $Al_2O_3$ are expected to be observed in the same pole figure.

Figure 7b shows well-defined spots at a tilt angle of about $\psi$ = 45°. Taking only into account the geometric considerations of simulated pole diagrams of $MoO_2$ (see supplementary material, S3), it is not possible to differentiate if the flakes have the (100) or the (001) crystalline plane parallel to the surface. This is because the spots in the two stereographic projections are located in very similar positions for both orientations. For example, assuming the surface plane is (100), the spots correspond to (110) and (1$\bar{1}$0) crystalline planes of $MoO_2$. Instead of two, six spots are observed because crystals are oriented in three equivalent in-plane directions onto the substrate surface, rotated by 120° relative to each other. If instead, it is assumed that the surface planes is (001), the spots would correspond to (011) and (0$\bar{1}$1) crystalline planes; also in this case with contributions of three different in-plane directions. These planes, belonging to the {011} and {110} families, form approximately 45° with both (100) and (001) surface plane orientations. For this reason, as commented above, the geometry of the spots distribution does not allow to determine the surface plane of $MoO_2$ flakes. However, intensity considerations do it: according the structure factor of the $MoO_2$ crystal, the {110} planes are expected to be prohibited while {011} ones are expected to exhibit the highest intensities. For this reason, it can be concluded that the present pole figure corresponds with the (001) $MoO_2$ plane parallel to the (001) $Al_2O_3$ substrate surface plane.

Around a tilt angle of 70°, very weak spots are observed (it was necessary to present the figure in logarithmic scale of intensities to appreciate these spots); they correspond to the low intensity (110), $(1\bar{1}0)$, $(\bar{1}\bar{1}1)$, $(\bar{1}11)$ reflections. They give place to twelve reflections taking into account the three different directions of the crystals, as in the previous case. The remaining peaks at around 57° correspond to the (0 1 2) $Al_2O_3$ at $2\theta$ angle of 25.6° (see stereographic projections in the Supplementary material, S3) rotated by 30° from the (110) planes of $MoO_2$. This is an evidence of textured epitaxial growth[15].

To determine the in-plane (epitaxial) relation of the $MoO_2/Al_2O_3$ heterostructure we used the above commented fact that, according to the pole figure, the azimuth of the (012) plane of $Al_2O_3$ is rotated by 30° with respect to the azimuth of the (110) plane of $MoO_2$. For $Al_2O_3$, the (012) plane shares azimuth with the (010) one, which is perpendicular to the (001) surface plane. On the other hand, in the $MoO_2$ crystal, the (110) plane shares azimuth with the (010) one, which is normal to the (001) substrate surface. Then, we can conclude that the (010) $Al_2O_3$ plane is rotated by 30° with respect to the (010) $MoO_2$ one. Considering that the (010) $Al_2O_3$ plane runs along the [100] direction and the (010) plane of $MoO_2$ runs along the [100] direction of $MoO_2$, it can be concluded that the flakes are oriented with the [100] direction of $MoO_2$ rotated by 30° with respect to the [100] direction of the sapphire substrate. It can be noted that this is just the epitaxial relation presented in Fig. 1 for configuration C.

Other authors have reported [7,8] that the epitaxial growth of $MoO_2$ onto (001) $Al_2O_3$ occurs with the (100) plane parallel to the (001) $Al_2O_3$ surface. But, as explained above, our pole figures correspond clearly with the (001) $MoO_2$ plane parallel to the substrate surface. On the other hand, as can be noted from our DFT calculations, the two orientations have similar strain energies, the (001) orientation being of slightly lower energy.

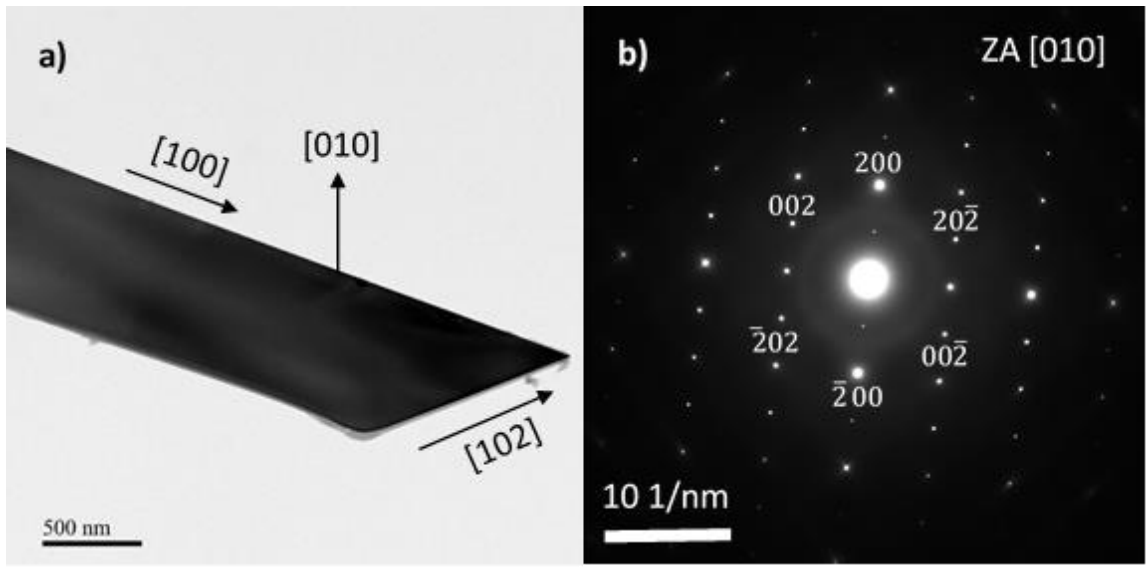

Fig. 8. a) TEM image of one of the standing flakes; b) selected area diffraction pattern showing the [010] zone axis

To complete the characterization of the standing flakes, TEM measurements were carried out. Fig. 8a shows an image of a standing flake that was taken off from the substrate. From this image, the height of the flakes can be estimated in the order of 1 µm. A selected area electron diffraction (SAED) pattern is displayed in Fig. 8b. This pattern can be simulated as corresponding with the [010] zone axis; this direction should be the normal to the flake's largest facet. On the other hand, we determined, from pole diagram analysis above, that the growth direction (the normal to the (001) plane) is the [102]. We can then conclude that the long axis of the flakes runs along the [100] direction of MoO₂ (or the [102] direction of the Al₂O₃ substrate surface). Since this direction coincides with that of the vertices of the lying flakes, it is very reasonable from the geometry of these flakes and their orientation with respect to the surface, that they are oriented in an epitaxial relation like that represented in the configuration B of Fig. 1. This configuration exhibiting the lowest strain energy according the DFT calculations supports this conclusion. The epitaxial relationships determined in each case are summarized in Table II.

Table II. Epitaxial relationship of the two kinds of MoO₂ flakes grown onto (001) oriented Al₂O₃.

| Standing flakes | $(001)MoO_2$ ∥ $(001)Al_2O_3$ and $[100]MoO_2$ ∥ $[120]Al_2O_3$ (configuration C) |
|---|---|
| Lying flakes | $(010)MoO_2$ ∥ $(001)Al_2O_3$ and $[100]MoO_2$ ∥ $[210]Al_2O_3$ (configuration B) |

5. Conclusions

Lying and standing flakes of MoO₂ were obtained onto (001)-oriented Al₂O₃ substrates using the simple and cost-effective CD-ICSVT technique. These flakes presented two different types of epitaxial relationships with the substrate whose origin was explored through DFT calculations of the strain energy involved in each growth pattern. The epitaxial relations as well as the orientation of the flakes with respect to the substrate were determined experimentally by means of SEM and TEM images and XRD and SAED patterns. It was found that the encountered epitaxial relations coincide with those of lower strain energy costs estimated by DFT. Besides proving that strain is the crucial factor determining the type of substrate/layer match in this heterostructure, this result validates our DFT theoretical approach for predicting strain-induced epitaxial relationships. This is also supported by the fact that flakes we have grown onto other surfaces, as A-plane Al$_2$O$_3$ or Si(100), show completely different morphologies and do not present epitaxial relationship with the substrate. (see supplementary material, S4). The presented procedure, which does not use excessive computer time or resources, should be useful in cases in which the epitaxial relationships are not obvious (like in the present case) and is expected to be more efficient than qualitative analysis of lattice parameters mismatches, which could afford ambiguous predictions. For example, in the present case, a qualitative analysis of both crystalline structures involved, would lead to hypothesize a different relationship in which $\vec{a}$ and $\vec{c}$ lattice vectors of MoO$_2$ forming an angle close to 120º to each other, should be oriented along $\vec{a}$ and $\vec{b}$ lattice vectors of Al$_2$O$_3$ (as in configuration A, see figure 1). However, after our DFT estimations, it was observed that this orientation implies a larger cost in terms of strain energy.

Standing flakes were found to be elongated with the [100] MoO$_2$ direction lying along the three equivalents [120] ones of Al$_2$O$_3$ surface. At late growth stage, standing flakes coalesce to form porous self-organized epitaxial films structured at nanometer scale. From the point of view of applications, this should be an interesting structure whenever large area and low defect material are demanded (like in solar cells, or catalysis, for example). Moreover, the porous structure will facilitate intercalation of different ions with advantages in application, ns for Li batteries electrodes or biosensing [16].

There are still puzzling aspects that cannot be completely understood with only strain considerations. For example, MoO$_2$ is not a lamellar material (like MoO$_3$ is) and the presence of flakes can only be explained by strong anisotropies in the growth rate of the different facets. To explain the geometry observed for the flakes it is necessary that the growth rate in the direction normal to the (010) plane (basal plane) is much smaller than in any other direction. It does not depend on the substrate. In fact, both, lying and standing flakes, which are differently matched to the substrates, have the same basal plane. Probably, the study of the morphology of the flakes in other kinds of substrates should bring clarity on these aspects. Some work is in progress in our team in this regard.

Acknowledgements. OdM and YG gratefully acknowledge the support from the program ''Cátedras de Excelencia de la Comunidad de Madrid (2016-T3/IND-1428)''. OdM and ER thanks the support of UNAM/DGAPA/PREI program 2018. The authors thanks A. Tavira and L. G. Pelayo for valuable technical assistance.